\documentclass[aip,reprint,floatfix,amsmath,amssymb,twocolumn]{revtex4-1}
\usepackage{graphicx}% Include figure files
\usepackage{dcolumn}% Align table columns on decimal point
\usepackage{bm}% bold math
\usepackage{epstopdf}
\usepackage{xcolor}
\usepackage{natbib}
\usepackage{subcaption} 
\usepackage{float}
\usepackage{tabularx}
\usepackage[a]{esvect}
\setcitestyle{square}

\begin{document}

\title{Leveraging extreme laser-driven magnetic fields for gamma-ray generation and pair production}

\author{O. Jansen}
\affiliation{Department of Mechanical and Aerospace Engineering, University of California at San Diego,\\La Jolla,
CA 92093, USA}
\affiliation{Center for Energy Research, University of California San Diego,\\ La Jolla, CA, 92093, USA}

\author{T. Wang}
\affiliation{Department of Mechanical and Aerospace Engineering, University of California at San Diego,\\La Jolla,
CA 92093, USA}
\affiliation{Center for Energy Research, University of California San Diego,\\ La Jolla, CA, 92093, USA}

\author{D. J. Stark}
\affiliation{Los Alamos National Laboratory, Los Alamos, NM 87545, USA}

\author{E. d'Humi\`eres}
\affiliation{Univ. Bordeaux-CNRS-CEA, Centre Lasers Intenses et Applications, UMR 5107, 33405 Talence, France}

\author{T. Toncian}
\affiliation{Institute for Radiation Physics, Helmholtz-Zentrum Dresden-Rossendorf e.V., 01328 Dresden, Germany}

\author{A. V. Arefiev}
\affiliation{Department of Mechanical and Aerospace Engineering, University of California at San Diego,\\La Jolla,
CA 92093, USA}
\affiliation{Center for Energy Research, University of California San Diego,\\ La Jolla, CA, 92093, USA}

\date{\today}

\begin{abstract}
%This work examines how plasma fields that are driven by an intense laser pulse in an over-critical plasma through relativistic transparency can be leveraged to enhance the rate of photon emissions. It is shown that a laser pulse of extreme intensity tends to drive a strong plasma magnetic field rather than a strong plasma electric field. This magnetic field stimulates emission by laser-irradiated electrons, causing them to generate a directed and dense beam of multi-MeV gamma-rays. Presented simulations show that more than $10^3$ electron-positron pairs can be created through a linear Breit-Wheeler process by colliding two such dense, collimated gamma-ray beams. The directivity of the gamma-rays beams makes it possible to achieve these numbers in a vacuum far away from the laser-irradiated targets that generate the gamma-rays.
The ability of an intense laser pulse to propagate in a classically over-critical plasma through the phenomenon of relativistic transparency is shown to facilitate the generation of strong plasma magnetic fields. Particle-in-cell simulations demonstrate that these fields significantly enhance the radiation rates of the laser-irradiated electrons, and furthermore they collimate the emission so that a directed and dense beam of multi-MeV gamma-rays is achievable. This capability can be exploited for electron-positron pair production via the linear Breit-Wheeler process by colliding two such dense beams. Presented simulations show that more than $10^3$ pairs can be produced in such a setup, and the directionality of the positrons can be controlled by the angle of incidence between the beams.

\end{abstract}

\maketitle

\section{Introduction}

New laser facilities, such as the Extreme Light Infrastructure (ELI)\cite{ELI2017} or multi-PW Apollon facility\cite{APOLLON.2017}, are expected to deliver laser pulses of unprecedented intensities~\cite{HABSELI}, making it possible to experimentally access the regime of laser-matter interactions with intensities approaching $10^{23}$ watt/cm$^2$. This prospect has generated a surge of interest in ultra-short laser-matter interactions because of the potential for fundamental physics research and radiation source applications\cite{Nature.4.130133} at these elevated field strengths. 

The key features of these interactions at extreme laser intensities are that electron dynamics in the target become ultra-relativistic and that quantum electrodynamic processes become relevant to the electron motion and energy distribution. 
Effectively, the laser-matter interaction in this case is more precisely a laser-plasma interaction, since the irradiated material becomes ionized long before the laser pulse reaches its peak intensity.
The relativistic energies of the electrons in these strong fields result in high radiative rates that play an integral role in altering the electron dynamics~\cite{PhysRevLett.113.134801}; as the laser intensity increases, so does the acceleration experienced by electrons in the irradiated plasma. This increase ultimately causes the classical picture of the emission and electron trajectories to break down, making the inclusion of quantum effects a necessity for the models~\cite{0741-3335-55-12-124018, PhysRevE.92.023305}. 

%~\cite{PhysRevLett.93.135005}

Previous work in this area of laser-plasma interactions at extreme intensities has shown that the efficiency of photon emissions can be surprisingly high due to the violent acceleration experienced by electrons in the target~\cite{PhysRevLett.108.165006,PhysRevLett.108.195001,Ji2014}. This acceleration can be induced not just by the laser pulse, but by plasma fields generated in the interaction as well~\cite{PhysRevLett.116.185003}. In this paper we examine how plasma fields that are driven by the irradiating laser pulse can be leveraged to enhance the rate of photon emissions, specifically showing that a laser pulse of extreme intensity tends to drive a strong plasma magnetic field rather than a strong plasma electric field. This magnetic field stimulates emission by laser-irradiated electrons, causing them to generate a directed and dense beam of multi-MeV gamma-rays.

Multiple applications specifically require beams of gamma-rays or can directly benefit from laser-driven gamma-ray sources~\cite{Nature.4.447}. Furthermore, these sources have the potential to open up new avenues of fundamental research that have been previously inaccessible in the laboratory. In particular, our understanding of the activity in the early Universe and high-energy astrophysics firmly relies on a so-called linear Breit-Wheeler process~\cite{PhysRev.46.1087}, in which a collision of two energetic photons creates an electron-positron pair. Even though this process has a significant impact on astrophysical phenomena~\cite{RevModPhys.76.1143,RUFFINI20101, NIKISHOV1962}, it has not been observed in laboratory conditions. Experiments thus far\cite{PhysRevLett.79.1626} have instead aimed at harnessing the non-linear Breit-Wheeler process, which utilizes more than two photons per collision. The difficulty in achieving the linear process stems from a low cross-section and a high energy threshold, which translates into needing a source that can supply dense beams of multi-MeV photons. We show that the laser-driven gamma-ray sources based on strong plasma magnetic fields can resolve this difficulty by delivering such a desired beam of photons. Our simulations indicate that a collision of two gamma-ray beams driven by a laser pulse with parameters similar to that of the Apollon facility in France can produce more than $10^3$ electron-positron pairs through the linear Breit-Wheeler process in a single shot even when the interaction region is significantly removed from the gamma-ray sources.

The rest of the paper is organized in the following way. In Sec.~\ref{Sec-2} we examine the dynamics of a single electron irradiated by a plane electromagnetic wave and estimate the threshold for a plasma magnetic field  above which the field significantly influences the photon emission by the electron. The fully self-consistent particle-in-cell simulations in Sec.~\ref{Sec-3} demonstrate that such a field can be achieved in laser-plasma interactions at extreme intensities by employing the effect of relativistically induced transparency. We calculate the resulting collimated photon beam and we then use two such beams in Sec.~\ref{Sec-4} to examine pair-creation via the linear Breit-Wheeler process. Finally, we summarize our findings and provide concluding remarks in Sec.~\ref{Sec-5}.

%*******************************************************

\section{Emission of a Single electron} \label{Sec-2}

It is known from classical electrodynamics that an electron accelerated by electric, $\bf{E}$, and magnetic, $\bf{B}$, fields emits electromagnetic radiation. The emitted power, $P$, is determined by the acceleration in an instantaneous rest frame. It is convenient to quantify this acceleration using a dimensionless parameter (\cite{LLclassicfields.1975,RIDGERS2014273} p.194)
\begin{equation} \label{eq:Eta}
\eta \equiv \frac{\gamma}{E_S} \sqrt{ \left({\bf E} + \frac{1}{c}\left[{\bf v}\times{\bf B}\right]\right)^2 - \frac{1}{c^2}\left({\bf E}\cdot{\bf v}\right)^2},
\end{equation}
where $\gamma$ and $\bf{v}$ are the relativistic factor and velocity of the electron, $c$ is the speed of light, and $E_S \approx 1.3 \times 10^{18}$ V/m is the Schwinger limit. The radiated power scales as $P \propto \eta^2$.

It is evident from Eq.~(\ref{eq:Eta}) that $\eta$, and thus the emission rate, increase with the amplitude of the fields acting on the electron. However, the field configuration and its orientation with respect to the electron velocity play a very important role as well\cite{JETP.35079708}. This becomes particularly apparent when considering an electron irradiated by a plane electromagnetic wave. In what follows, we consider several simple yet insightful examples with an increasing degree of sophistication.

In all of the examples, we consider a plane linearly-polarized electromagnetic wave with wavelength $\lambda$ propagating along the $x$-axis. Without any loss of generality, we assume that ${\bf{E}}_{wave} = E {\bf{e}}_y$ and ${\bf{B}}_{wave} = B {\bf{e}}_z$, where ${\bf{e}}_y$ and ${\bf{e}}_z$ are unit vectors. Here we invoke the well-known result that $E = B$ for a wave propagating in the positive direction along the $x$-axis.

%The wave electric and magnetic fields are then only a function of a normalized phase
%\begin{equation}
%\xi \equiv \frac{2 \pi}{\lambda} \left(x - ct \right), 
%\end{equation}
%where $t$ is the time in the laboratory frame of reference. 

Our first example is an ultra-relativistic electron colliding head-on with the wave without any appreciable transverse motion, such that ${\bf v} = -v {\bf{e}}_x$. It readily follows from Eq.~(\ref{eq:Eta}) that
\begin{equation} \label{eq:Eta2}
	\eta_{counter} \approx 2\gamma \frac{E}{E_S},
\end{equation}
where $\gamma = 1 / \sqrt{1 - v^2}$. This result is what one might expect based on the structure of the expression for the parameter $\eta$, where both the field amplitude and the $\gamma$-factor enter the numerator. This notably permits considerable enhancement of the emission rate due to the relativistic factor, which we will observe is not the case for alternate orientations.

Let us now consider an ultra-relativistic electron that is co-moving with the wave, so that ${\bf v} = v {\bf{e}}_x$. It follows from Eq.~(\ref{eq:Eta}) that the value of $\eta$ is now greatly reduced:
\begin{equation}
\label{eq:Eta1}
	\eta_{co} \approx \frac{1}{2\gamma} \frac{E}{E_S},
\end{equation}
An important aspect here is that $\eta$ is inversely proportional to the $\gamma$-factor. This is because for a co-moving ultra-relativistic electron the acceleration induced by the electric field of the wave is strongly compensated by the acceleration induced by the magnetic field of the wave. The reduction of the parameter $\eta$ indicates that the emitted power is strongly suppressed in this case.

These two examples clearly articulate why colliding relativistic electrons with an intense laser pulse is one of the more promising avenues of generating intense radiation and even of examining the effects of radiation reaction~\cite{PhysRevLett.113.134801,PHUOC101038}. However, it is worth pointing out that colliding setups typically require two laser pulses, and this severely constrains most facilities' capabilities of employing this technique. An additional laser pulse is needed to accelerate electrons and generate an ultra-relativistic bunch for the collision. 

In contrast to the colliding setup, the co-propagating setup can be achieved using just a single laser pulse. This is because an intense laser pulse can accelerate an electron in the direction of the pulse propagation through so-called direct laser acceleration. If the normalized laser amplitude, $a_0 \equiv |e| E_0 / m_e c \omega$, is large, $a_0 \gg 1$, then the laser field induces relativistic electron motion which in turn leads to preferentially forward acceleration. Here $m_e$ and $e$ are the electron mass and charge, while $E_0$ and $\omega$ are the pulse electric field amplitude and its frequency. However, transverse electron oscillations are essential for this mechanism of acceleration, and our previous estimates neglected such oscillations and the resulting increase of the longitudinal momentum that can be substantial~\cite{PhysRevLett.111.065002,arefiev_robinson_khudik_2015,PhyoPlas056704}, essentially implicitly assuming that $a_0 \ll 1$. We thus have to revise the estimates for $\eta$ by taking transverse oscillations into account.

%In these examples, we have neglected the transverse electron motion, and transverse oscillations induced by the wave are clearly inconsequential for the counter-propagating case. However, they need to be carefully accounted for in the co-propagating regime, because the observed compensation in the electron acceleration is a direct result of the assumption that the electron is only moving forward. Another reason why the co-propagating case requires extra care is that an intense laser pulse can substantially increase the electron energy, thus changing the $\gamma$-factor through so-called direct laser acceleration, which we briefly overview next.

We now briefly overview the direct laser acceleration mechanism in order to provide the context for the result that will follow. Electron acceleration by an intense laser pulse can be conveniently described using the normalized vector potential ${\bf{a}}$. This potential specifies the laser fields:
\begin{eqnarray}
&& {\bf{E}}_{wave} = -\frac{m_e c}{|e|} \frac{\partial {\bf{a}}}{\partial t}, \\
&& {\bf{B}}_{wave} = \frac{m_e c^2}{|e|} \nabla \times {\bf{a}},
\end{eqnarray}
where $m_e$ and $e$ are the electron mass and charge respectively. In our case the potential is only a function of a normalized phase
\begin{equation}
\xi \equiv \frac{2 \pi}{\lambda} \left(x - ct \right),
\end{equation}
so that ${\bf a} = a(\xi) {\bf{e}}_y$. For simplicity, let us assume that the amplitude gradually increases and then remains constant, such that $a(\xi) = a_0 \sin(\xi)$. 
%As already mentioned, the amplitude $a_0$ is related to the amplitude of the electric field $E_0$ via the relation $a_0 = |e| E_0 / m_e c \omega$.
The momentum components of an initially static electron that is irradiated by such a wave are then prescribed by\cite{PhyoPlas056704}
\begin{eqnarray}
	&& p_x / m_e c = a^2 / 2, \label{EQ-px}\\
	&& p_y / m_e c = a. \label{EQ-py}
\end{eqnarray}
We see that in addition to oscillating with the electric field, the electron also moves forward along the propagation direction. The electron motion is strongly relativistic for wave amplitudes $a_0 \gg 1$, and that is why such amplitudes are called relativistic. The $\gamma$-factor strongly oscillates at these amplitudes,
\begin{equation} \label{EQ-gamma}
\gamma = 1 + a^2 /2,
\end{equation}
and the electron motion is predominantly directed forward with an angle to the $x$-axis that can be estimated as $p_y/p_x \sim 1/a$.

We are now well-equipped to calculate the normalized acceleration $\eta$ for the electron that undergoes the direct laser acceleration at relativistic wave amplitudes. It follows directly from Eq.~(\ref{eq:Eta}) and Eqs.~(\ref{EQ-px}) and (\ref{EQ-py}) for the electron momentum that
\begin{equation} \label{eta-DLA}
	\eta = \frac{E}{E_S} \left( \gamma - \frac{p_x}{m_e c} \right)\rightarrow \eta_{DLA} = \frac{E}{E_S}.
\end{equation}
The left equality assumes nothing about the transverse electron oscillations and it can thus be used to calculate $\eta$ both for the already considered co-propagating case without the transverse motion and for the electrons accelerated using the direct laser acceleration mechanism (see right equality). By only assuming that $|p_x| \gg |p_y|$, we find that for an ultrarelativistic electron
\begin{equation}
\gamma - \frac{p_x}{m_e c} \approx \frac{1}{2 \gamma} \left( 1 + \frac{p_y^2}{m_e^2 c^2} \right).
\end{equation}
This result directly shows how even a relatively small but relativistic transverse momentum can significantly enhance the value of $\eta$, which explains the difference between $\eta_{DLA}$ and $\eta_{co}$ given by Eqs.~(\ref{eta-DLA}) and (\ref{eq:Eta1}).

\begin{figure} [H]
   \begin{center}
      \includegraphics[width=\columnwidth] {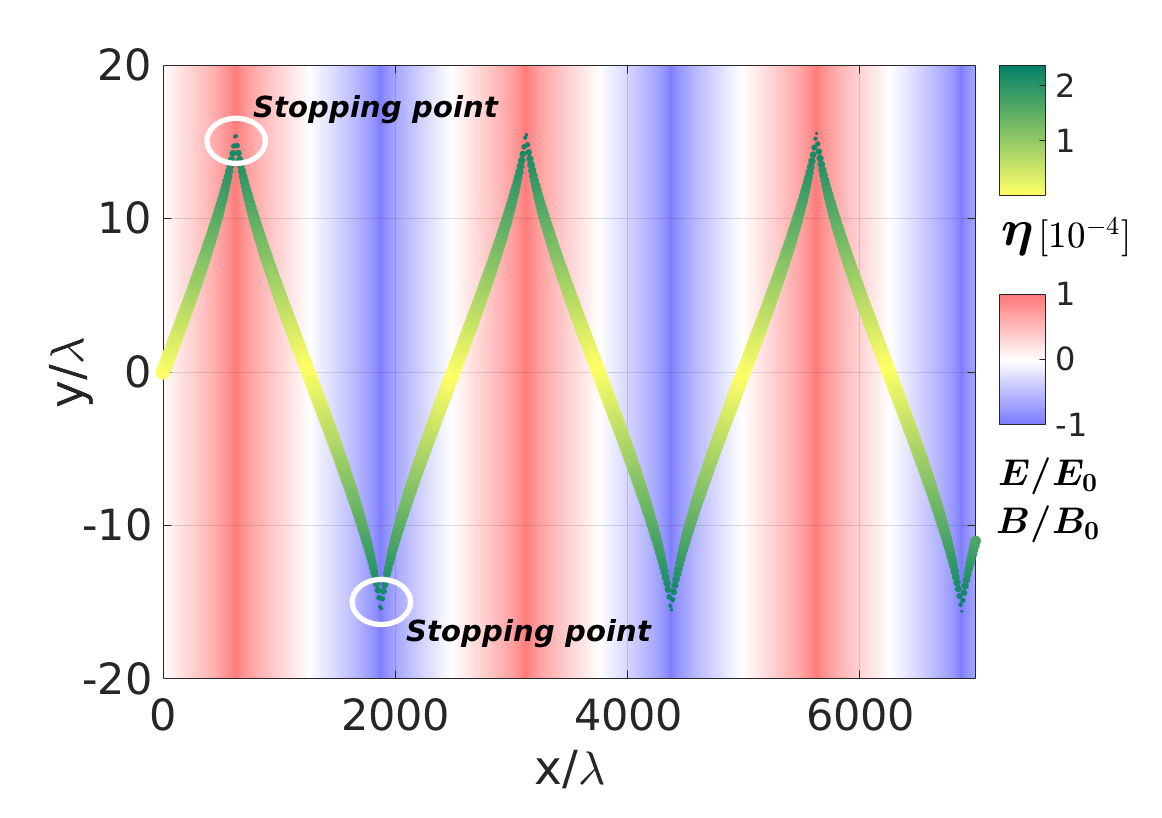}
      \caption{\label{fig:EtaVacuum} Electron trajectory in a plane wave with a normalized amplitude $a_0 = 100$. The background color represents the wave electric and magnetic fields acting on the electron, normalized to their maximum amplitudes. The relative size of the markers along the trajectory represents the changing $\gamma$-factor, while the color-coding represents the value of $\eta$.}
   \end{center}
\end{figure}

Despite the substantial increase of $\eta$, and thus the emission, in the direct laser acceleration regime, compared to the co-propagating case, the emitted power is still significantly lower than in the case where the electron collides with the laser pulse. In order to gain further insight and understand what limits the emission during the direct laser acceleration, we examine a corresponding electron trajectory. Figure~\ref{fig:EtaVacuum} provides comprehensive information about an electron accelerated from rest by a laser pulse that is gradually ramped up in amplitude to $a_0 = 100$ by showing both the electric and magnetic fields acting on the electron (after the pulse reaches its peak) as well as the electron $\eta$ and its relativistic $\gamma$-factor. One striking feature is that the electron trajectory has stopping points when the laser fields reach their peak amplitude. It then immediately follows from Eq.~(\ref{eta-DLA}) that the parameter $\eta$ also peaks in the vicinity of the stopping points. At these points, the electron halts its motion and its $\gamma$-factor drops. We therefore arrive at an important conclusion that most of the radiated emission during the direct laser acceleration occurs when the electron energy is relatively low.

The observed phase shift or phase mismatch between the $\gamma$-factor and the laser electric field $E$ is characteristic of the direct laser acceleration. It necessarily reduces the value of $\eta$, as evident from Eq.~(\ref{eq:Eta}). Since the laser electric and magnetic fields drop as the electron reaches its maximum energy, the energy increase cannot be efficiently utilized to increase the emitted power. Moreover, the counter-synchronism also suppresses emission of high energy photons. The maximum energy of emitted photons is limited by the electron energy, so the photons emitted near the turning points where most of the emissions occur would have the lowest energy of all the emitted photons.

While the counter-synchronism is a major downside of the direct laser acceleration, it can be successfully negated by introducing quasi-static plasma fields. At this stage, let us assume that the forward propagating laser pulse drives a longitudinal electron current that sustains a slowly-evolving azimuthal magnetic field coiled around the axis of the laser pulse. The details of the magnetic field generation will be carefully examined in the following section. In order to provide a simple estimate for the effect of this field on $\eta$, we set ${\bf{E}} = E {\bf{e}}_y$ and ${\bf{B}} = (B + B_*) {\bf{e}}_z$, where $E$ and $B$ are the fields of a plane wave, while $B_*$ is the static plasma magnetic field. 

We now assume that the plasma field $B_*$ is so weak that its effect on the electron trajectory in the direct laser acceleration regime is negligible. We are looking for the threshold value of $B_*$ where it begins to affect our estimates for $\eta_{DLA}$. Using expressions~(\ref{EQ-px}) - (\ref{EQ-gamma}) for the electron momentum and its $\gamma$-factor, we find that 
\begin{equation} \label{EQ-12}
\eta \approx \frac{E}{E_S} \sqrt{1 + \kappa + \kappa^2},
\end{equation}
where for compactness we introduce a dimensionless quantity
\begin{equation}
\kappa \equiv \gamma B_* / E.
\end{equation}
If $|\kappa|$ is small, then the expression (\ref{EQ-12}) reduces to $\eta \approx \eta_{DLA}$. However, the expression for $\eta$ changes as $|\kappa|$ approaches unity. We can therefore set $|\kappa| \approx 1$ as the defining criterion for critical value $B_{*cr}$. It is notable that according to these estimates the magnetic field of the plasma should increase $\eta$ --- and, as a result, the emission --- even at amplitudes that are still well below the amplitude of the laser magnetic field:
\begin{equation}
B_{*cr} \approx B / a_0^2 \ll B.
\end{equation}
Here we once again take into account that in a plane wave the amplitudes of the electric and magnetic fields are the same, $E = B$, and that the $\gamma$-factor  at highly-relativistic laser amplitudes scales approximately as $\gamma \approx a_0^2$.

Our estimates clearly indicate that even a plasma magnetic field of relatively moderate strength can significantly boost the photon emission during an electron's direct laser acceleration. If the required magnetic field can be generated by driving an intense laser pulse through a plasma, then this would make this concept a very promising method for creating significant numbers of high-energy photons with just a single laser pulse. Note that we assumed the energy gain by the electron remains unaffected by this plasma field, but it is plausible that this magnetic field can enhance the energy gain by changing the frequency of the transverse momentum oscillations. A somewhat similar effect has been observed for relatively weak transverse static plasma electric fields in laser-generated channels~\cite{Phy_Plas103108,PhysRevLett.108.145004}, albeit at significantly lower intensities than the ones we are considering here.

%*******************************************************

\section{Emission enhanced by plasma magnetic fields} \label{Sec-3}

The estimates presented in the previous section indicate that magnetic fields driven in a plasma can prove advantageous for the photon emission from laser-accelerated electrons. Motivated by these estimates, we here examine this scenario using fully self-consistent particle-in-cell simulations in order to characterize the generated photon beam when using a laser pulse with parameters similar to the constructed Apollon laser in France.

\begin{figure} [H]
   \begin{center}
      \includegraphics[width=\columnwidth]{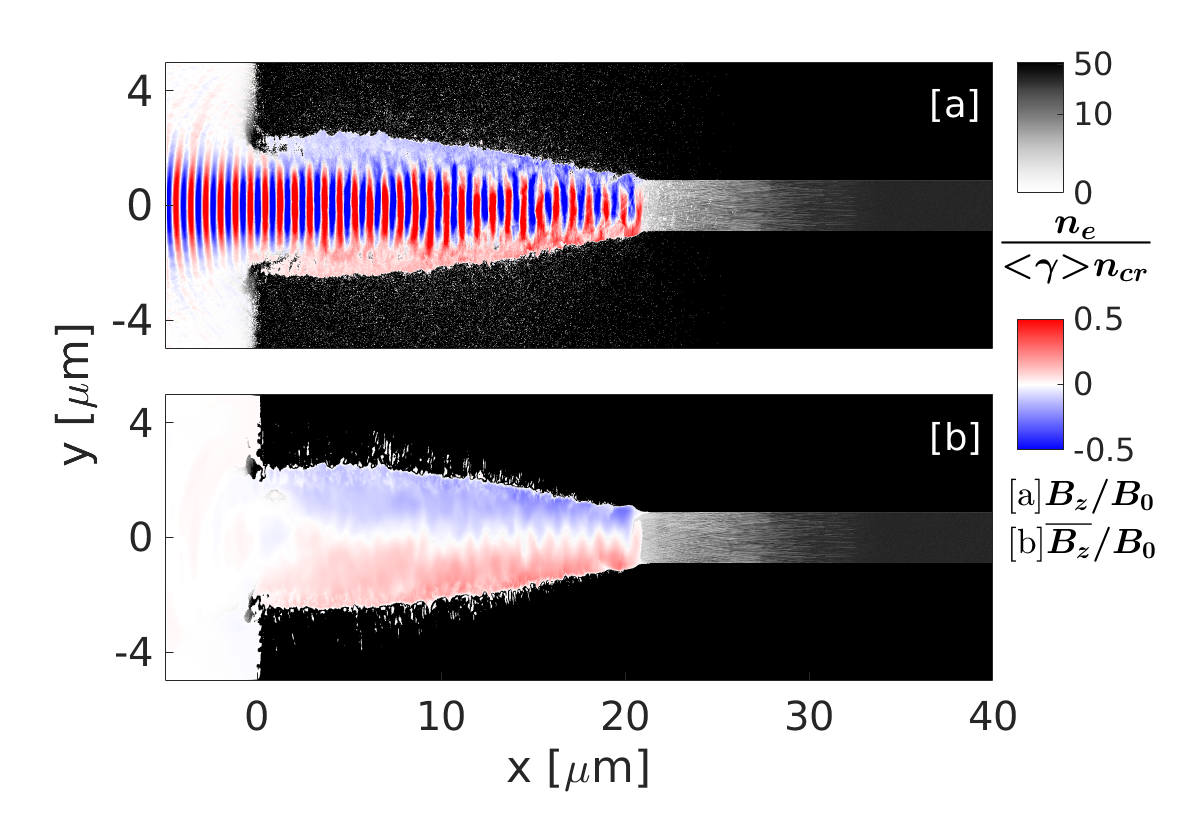}
      \caption{\label{fig:ChannelMovIons} Snapshots taken at $t \approx 27$ fs of a structured target irradiated by a laser pulse with $a_0 \approx 257$. The gray-scale shows the relativistically adjusted electron density. The red and blue color-scale shows the $z$ component of the magnetic field, the total field in the upper panel and the time-averaged field (over two laser cycles) in the lower panel.}
      \end{center}
\end{figure}

It is intuitively clear that using a dense plasma would be beneficial for driving strong magnetic fields, because the current that sustains the field scales linearly with the plasma electron density. The laser can only drive such a current, though, as long as it can propagate through the plasma. At non-relativistic laser amplitudes, $a_0 \ll 1$, the cutoff electron density above which the plasma becomes opaque is determined only by the frequency of the incoming laser pulse through $n_{cr}=m\omega^2/(4\pi e^2)$. This is often referred to as the classical critical density.

In contrast, a high-intensity laser pulse with $a_0 \gg 1$ can make an otherwise opaque plasma with a classically over-critical electron density $n_e \gg n_{cr}$ transparent. As the laser electric field accelerates plasma electrons to relativistic energies, it changes the optical properties of the plasma. A simplistic view is to treat this as an effective mass increase by a factor of $\gamma$, where $\gamma$ is the characteristic relativistic factor of the electron population. Even though this qualitative analogy to predict a $\gamma$ factor enhancement to the critical density is helpful, it should be pointed out that it has limited applicability~\cite{PhysRevLett.115.025002}. The effect described above  has been termed as the relativistically induced transparency. 

This phenomenon of relativistic transparency offers an attractive possibility of driving very strong plasma currents in classically over-critical plasmas with intense laser pulses.  The plasma becomes opaque at $n_e / \gamma \approx n_{cr}$ and, taking into account that the $\gamma$-factor is driven by the laser to $\gamma \approx a_0$, the relativistically adjusted critical density $n_*$ can be estimated as $n_* \approx a_0 n_{cr}$. Driving plasma currents in plasmas with $n_e$ close to the relativistically adjusted critical density, however, presents a challenge, because the laser pulse propagation becomes extremely unstable.

\begin{figure} [H]
   \begin{center}
      \includegraphics[width=\columnwidth]{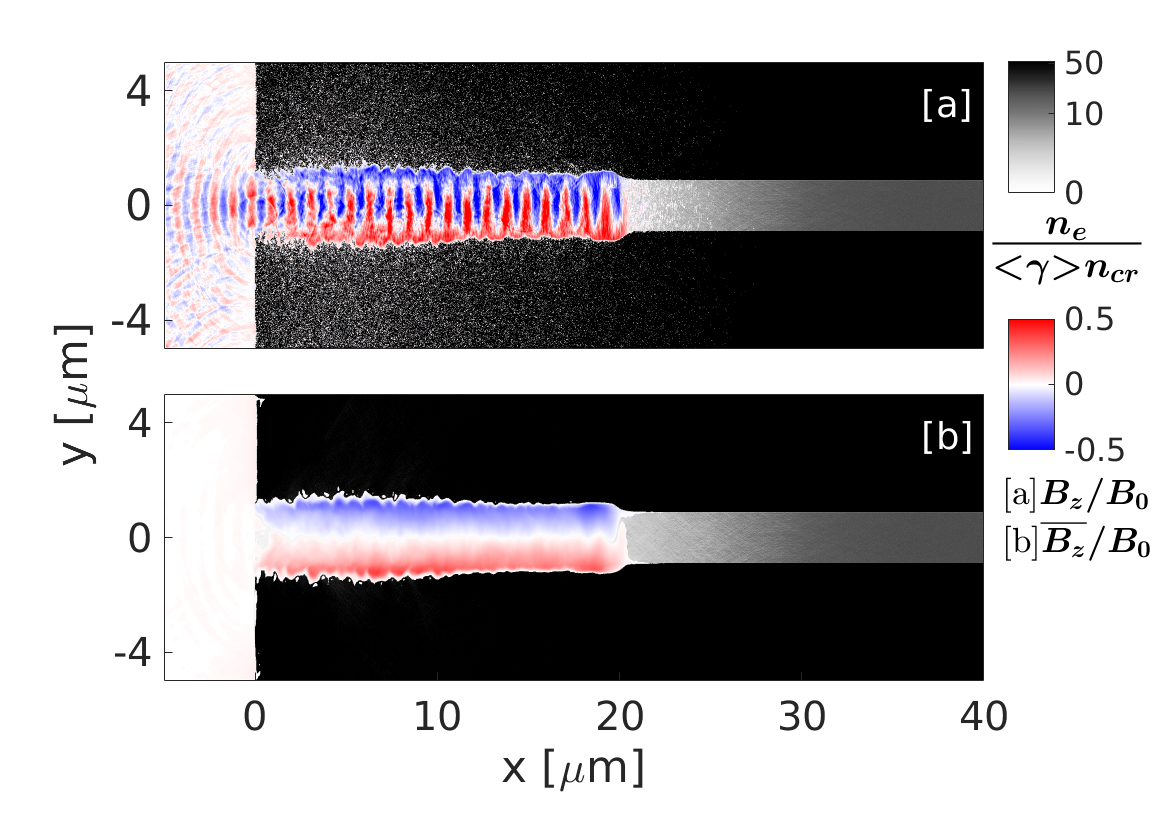}
      \caption{\label{fig:ChannelStatIons} Snapshots taken at $t \approx 27$ fs of a structured target with immobile ions irradiated by a laser pulse with $a_0 \approx 257$. The gray-scale shows the relativistically adjusted electron density. The red and blue color-scale shows the $z$ component of the magnetic field, the total field in the upper panel and the time-averaged field (over two laser cycles) in the lower panel.}
      \end{center}
\end{figure}

Structured targets allow one to overcome the laser stability issues while taking full advantage of working with $n_e \gg n_{cr}$ to generate strong magnetic fields~\cite{PhysRevLett.116.185003}. The basic idea is to use a target with a channel that becomes relativistically transparent when irradiated by the laser pulse. The bulk of the target has a higher electron density than that in the channel, which allows for optical guiding of the laser pulse. The upper panel in Fig.~\ref{fig:ChannelMovIons} illustrates such guiding in a two-dimensional PIC simulation. 

In the simulation whose results are shown in Fig.~\ref{fig:ChannelMovIons}, we use a linearly-polarized 800 nm laser pulse with a peak intensity of $7 \times 10^{22}$ W/cm$^2$, as mentioned before, similar to the Apollon facility. The pulse propagates in the positive direction along the $x$-axis with its electric field polarized in the plane of the simulation [the $(x,y)$-plane]. The pulse is 90 fs long and has a focal spot of 1.1 $\mu$m (full-width at half maximum in regard to the intensity), focusing at normal incidence onto the entrance of the channel.

For simplicity, we use a target that consists only of electrons and protons. The bulk target density is $n_e = 100 n_{cr}$, which makes it relativistically near-critical for the considered laser pulse with a peak normalized intensity of $a_0 \approx 257$. The channel electron density is set at $n_e = 10 n_{cr}$. This choice was guided by target manufacturing considerations~\footnote{Private communication with Dr. Mingsheng Wei and Dr. Jarrod Williams at General Atomics}, as availability of materials and techniques required for the considered target are critical for the implementation of the discussed concept. The initial radius of the channel is $R = 0.9$ $\mu$m to provide good coupling of the laser energy into the channel. We use 20 macro-particles per cell to represent electrons and 20 macro-particles per cell to represent protons. The spatial resolution is 50 cells per $\mu$m along the $x$-axis and 100 cells per $\mu$m along the $y$-axis. The simulation is performed using EPOCH\cite{Epoch}.

\begin{figure} [H]
   \begin{center}
      \includegraphics[width=\columnwidth,trim={0.2cm 0.5cm 0 0.4cm},clip]{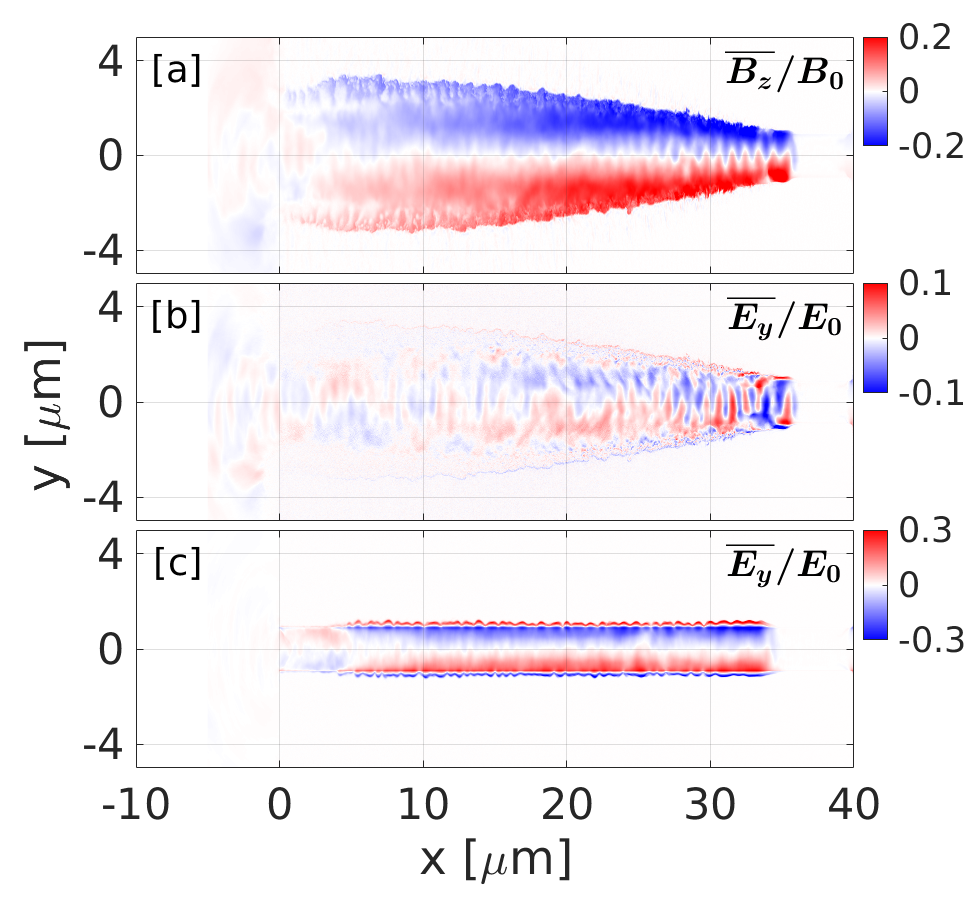}
      \caption{\label{fig:EBcomparison} Snapshots of time-averaged electric and magnetic fields taken at $t \approx 107$ fs. The top two panels correspond to the simulation with mobile ions, while the lower panel corresponds to the simulation with immobile ions. }
      \end{center}
\end{figure}

The change in optical properties of the plasma induced by the laser pulse is depicted in Fig.~\ref{fig:ChannelMovIons} by plotting the normalized relativistically adjusted electron density $n_e / \gamma n_{cr}$, where $\gamma$ is the cell-averaged electron relativistic $\gamma$-factor. The channel becomes ``lighter'' in the presence of the laser pulse, which signifies that it is optically transparent. The snapshots are taken at $t \approx 27$ fs, with $t = 0$ fs defined as the time when the laser pulse would reach its peak amplitude in the focal plane at $x = 0$ $\mu$m in the absence of the target. 

In line with our expectations, the laser pulse drives a strong slowly-evolving magnetic field as it propagates along the relativistically transparent channel. As evident from the lower panel of Fig.~\ref{fig:ChannelMovIons}, the amplitude of the time-averaged magnetic field is comparable to the instantaneous values dictated by the oscillating laser field. The averaging is performed over two laser cycles, and the field is normalized to $B_0 \approx 1.7$ MT, which is the peak amplitude of the magnetic field in the focal plane at $x = 0$ $\mu$m in the absence of the target. For reference, the similarly defined peak amplitude of the  electric field is $E_0 \approx 5.13 \times 10^{14}$ V/m.

\begin{figure} [H]
   \begin{center}
      \includegraphics[width=\columnwidth]{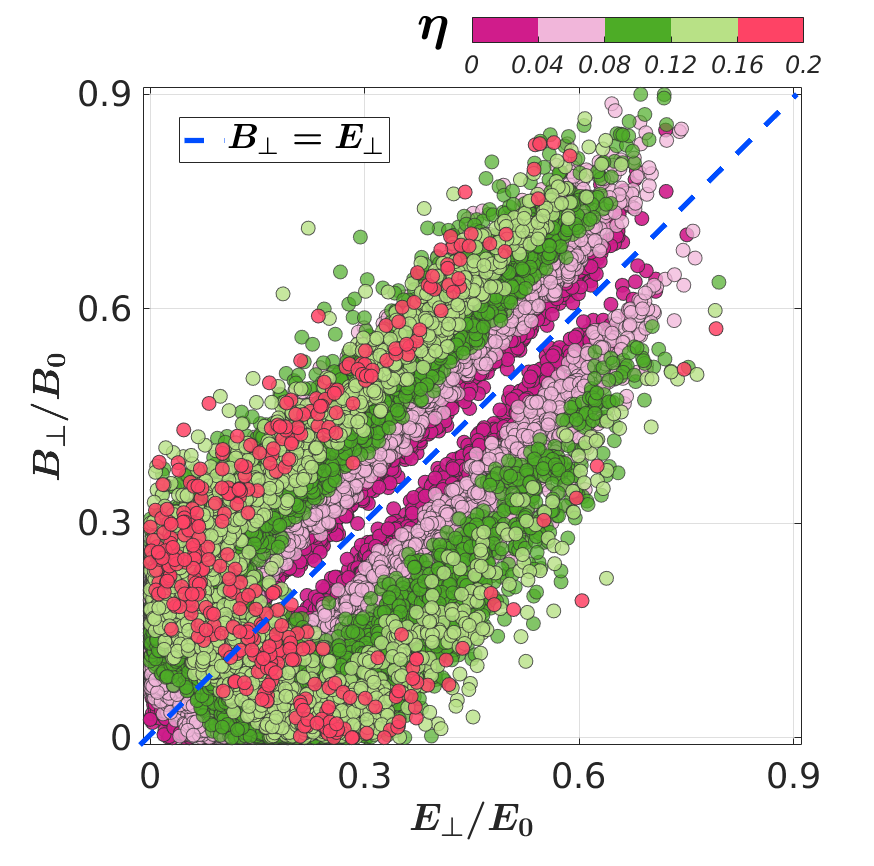}
      \caption{\label{fig:EtaOfFields} 
      Scatterplot of emitting electrons in terms of the transverse electric and magnetic fields that they experience at the timestep of emission. Each circle represents an emission event, and they are color-coded by the characterizing $\eta$ value. The data set was downsized by only selecting electrons with energies above 250 MeV that emit photons with energies above 20 MeV.
      }
      \end{center}
\end{figure}

Also of note from this simulation is that the channel in Fig.~\ref{fig:ChannelMovIons} is significantly wider than it was prior to being irradiated by the pulse. This clearly indicates that there is non-negligible transverse ion movement, which then raises a question of the role that the ion mobility plays in the considered setup. In order to gain more insight, we have performed another simulation where the ions are immobile while all other parameters remain unchanged. Snapshots of the instantaneous and time-averaged magnetic fields from this simulation are shown in Fig.~\ref{fig:ChannelStatIons}. Taking transverse lineouts at $x = 10$ $\mu$m, we find from Figs.~\ref{fig:ChannelStatIons} and \ref{fig:ChannelMovIons} that the peak value of the time-averaged magnetic field decreases from roughly 0.75 MT to 0.5 MT by switching from immobile to mobile ions. The reduction in the case of mobile ions is possibly caused by the electron current being distributed over a larger channel cross-section. Nevertheless, the quasi-static magnetic field remains very strong and it has an unprecedented amplitude that is roughly 25\% of the oscillating magnetic field in the laser pulse.

We can therefore conclude that the generation of a strong quasi-static magnetic field is a robust process with respect to the ion mobility. The situation is very different, however, for quasi-static transverse electric fields that are also generated in the channel due to charge separation caused by the laser pulse. Figure~\ref{fig:EBcomparison} shows snapshots of time-averaged electric fields from the simulations with mobile (middle panel) and immobile (lower panel) ions. Clearly, the ion mobility dramatically reduces the electric field, as its peak value drops by more than an order of magnitude from $E_y/E_0 \approx 0.3$ to $E_y/E_0 \approx 0.03$ in the cross-section at $x = 20$ $\mu$m.

It must be emphasized that in the case of immobile ions the relative strengths of the time-averaged electric and magnetic fields are comparable: $B_z / B_0 \sim E_y / E_0$. In contrast to that, we have $B_z / B_0 \gg E_y / E_0$ if the ions are mobile. This feature is clearly noticeable when comparing the top two panels in Fig.~\ref{fig:EBcomparison}. This result justifies our earlier focus on the slowly-evolving plasma magnetic field as the primary mechanism for enhancing the electron emission. From here onwards we restrict our discussion to the simulation with mobile ions.

\begin{figure} [H]
   \begin{center}
      \includegraphics[width=\columnwidth]{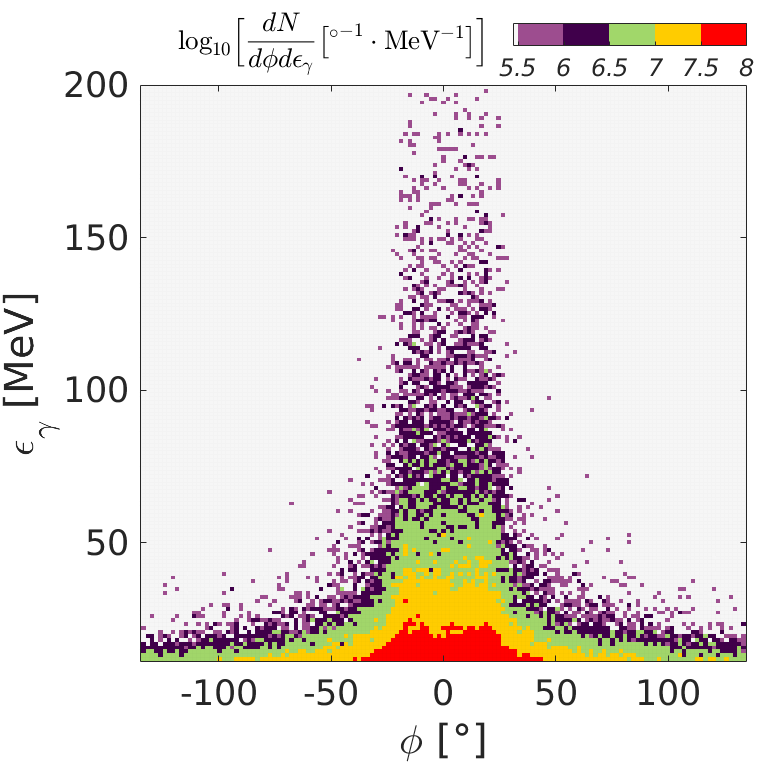}
      \caption{\label{fig:PhotAnglHist} Angular distribution of the photons emitted in a structured target irradiated by a laser pulse with $a_0 \approx 257$ as shown in Fig.~\ref{fig:ChannelMovIons}. These are all of the emissions that have occurred by  $t \approx 107$ fs. A large number of high energy photons is emitted forward into an opening angle of $\pm 30^\circ$.}
      \end{center}
\end{figure}

A direct way to quantify the impact of the plasma magnetic field on the emission is to examine the value of the parameter $\eta$ for emitting electrons. The emission process in our simulation is a discrete process where individual photons are probabilistically emitted by accelerated electrons. The corresponding algorithm implemented in EPOCH has been discussed in detail in Ref.~\cite{0741-3335-53-1-015009}. We have recorded the data for all of the photon emissions during the simulation shown in Fig.~\ref{fig:ChannelMovIons} for post-processing, including the electric and magnetic fields acting on each electron during the timestep of emission. 

We are primarily interested in high-energy electrons with $\gamma \gg a_0$ due to their potential to radiate multi-MeV photons. During the direct laser acceleration in a vacuum, the value of $\eta$ drops with the increase of the $\gamma$-factor because of the counter-synchronism in oscillations of $\gamma$ and of the laser electric field. In other words, high-energy electron emission is deemed to be inefficient in the absence of plasma fields (see Sec.~\ref{Sec-2}). Motivated by this observation, we now isolate the emissions by electrons with energies above 250 MeV during our simulation. We further restrict the dataset to photon emissions with energies above 20 MeV. Remarkably, the parameter $\eta$ reaches values as high as 0.2 (see Figure~\ref{fig:EtaOfFields}), which validates our conjecture that the emission by energetic electrons becomes significantly more efficient in the presence of the plasma magnetic field.

The profound difference between the emission in a plasma channel and in a vacuum becomes more apparent when examining the local fields experienced by the selected electrons at their moments of emission. In Figure \ref{fig:EtaOfFields} we present a scatterplot showing the perpendicular (to its instanenous momentum) electric and magnetic fields for each electron when emitting, each value color-coded by the corresponding $\eta$ of the particle. Here we observe that a significant number of emissions occur where there is a discrepancy between the local $E_\perp$ and $B_\perp$ field strengths. 
%The distance to which the emissions strafe from the central axis is about $1/3B_0$, which also is the strength of the field generated by the channel. For the 2D case, where $x$ is the propagation direction of the laser $E_\perp = E_y$ and $B_\perp = B_z$. 
It is important to point out that the emissions further away from the central axis $E_\perp/E_0 = B_\perp/B_0$ show a higher value of $\eta$ than the emissions on-axis where particles oscillating in vacuum would reside. These highest $\eta$ values can be found at an offset of about $B_\perp/(3B_0)$ from the central diagonal, which is roughly the strength of the magnetic field at the edge of the channel. Therefore, the strong plasma magnetic field generated by the laser-driven current can account for the elevated emission rates of the ultra-relativistic electrons that are observed in the simulations. 

%This is also where the emission starts to break off.

%Since the electrons are accelerated by the laser in the forward direction through direct laser acceleration, this directly translates into a directed beam of photons.

This enhancement of the emission from high energy electrons in strong magnetic fields has the additional benefit of corresponding with strongly collimated emission. Since the opening cone of the emission angle is inversely proportional to the Lorentz-factor of the emitting electron, $\Delta \alpha \sim 1/\gamma$, the ultra-relativistic electrons will essentially emit parallel to their trajectories. Figure \ref{fig:PhotAnglHist} plots the angular distribution (with respect to the laser propagation direction) of the emitted photons across the observed energy range, showing that the emission of the highest energy photons is limited to a range of approximately $\pm 30^\circ$. These photon yields are calculated by assuming that the generated beam of photons from the 2D simulation has a transverse (out-of-plane in the $z$-direction) dimension of 1 $\mu$m, comparable with the focal spot size. As the magnetic field can tightly constrain the electron motion in the channel~\cite{PhysRevLett.116.185003}, this narrow angular spread is expected. It is this combination of the enhanced gamma-ray production and the high degree of collimation, both facilitated by the quasi-static magnetic field, that opens up exciting avenues of application for gamma-ray sources. In the next section we demonstrate how this technique can be exploited for one such prospect.

%*******************************************************

\section{Pair production via Photon-Photon Collisions} \label{Sec-4}

\begin{figure} [H]
   \begin{center}
      \includegraphics[width=0.6\columnwidth]{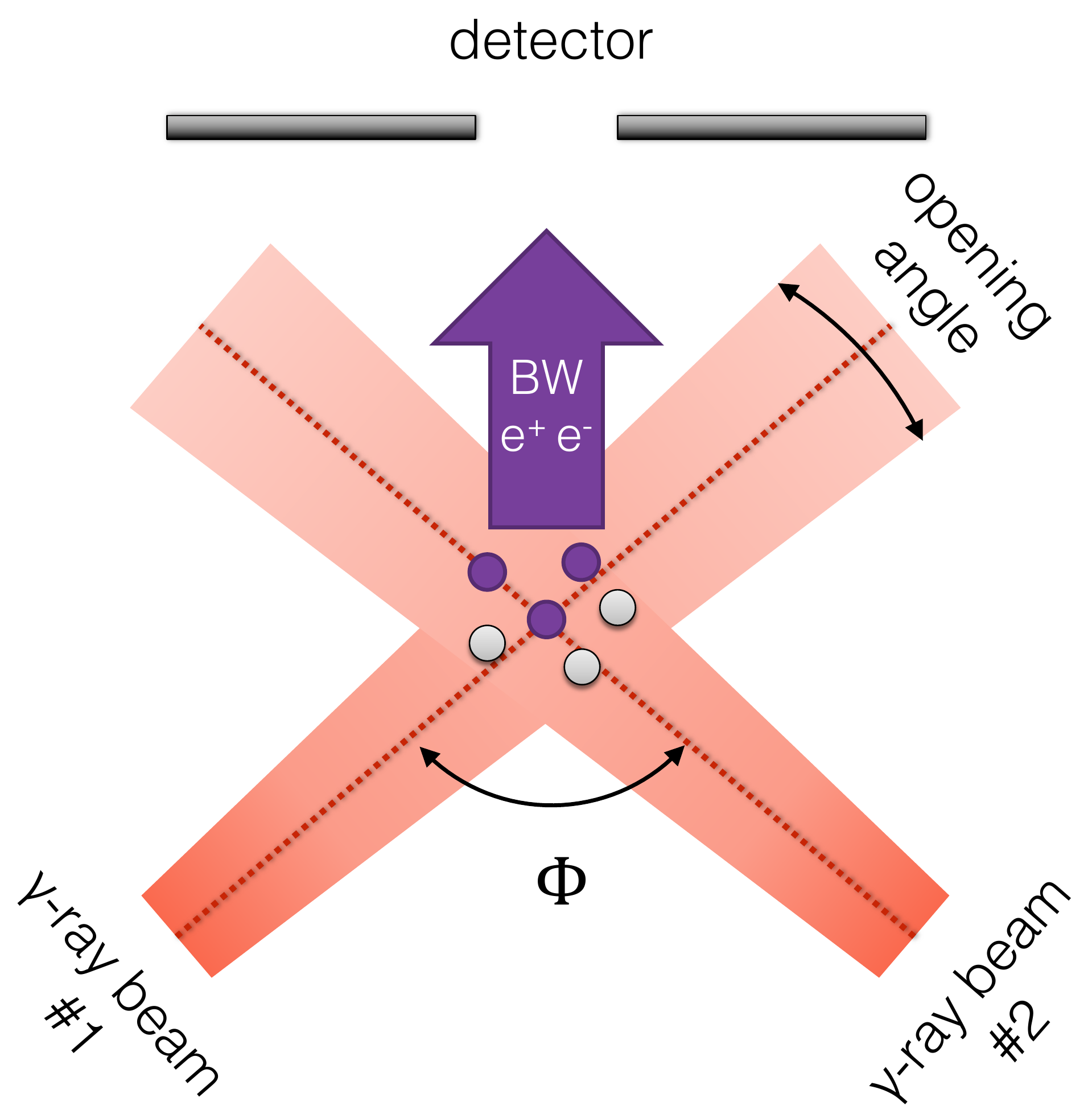}
      \caption{\label{fig:BWexperiment} Schematic setup for pair production via two-photon collisions by employing two $\gamma$-ray beams.}
      \end{center}
\end{figure}

In Sec.~\ref{Sec-3}, we demonstrated how relativistic transparency and strong quasi-static magnetic fields can be leveraged to generate a well-directed beam of energetic photons (gamma-rays). The directivity of the beam opens up the exciting possibility of colliding two such beams away from the target in a vacuum to generate electron-positron pairs through a linear Breit-Wheeler process~\cite{PhysRev.46.1087}. The corresponding setup is schematically shown in Fig.~\ref{fig:BWexperiment}.

As already mentioned in the introduction, the linear Breit-Wheeler process plays a fundamental role in astrophysical phenomena, but it has not yet been directly observed  in laboratory conditions.  The cross-section for pair production by two photons with energies $\epsilon_{\gamma_1}$ and $\epsilon_{\gamma_2}$ colliding at an angle $\Phi$ is given by~\cite{BERESTESKII.1982}
\begin{equation} \label{BWcross}
	\sigma_{\gamma \gamma} = \frac{\pi}{2}r_e^2(1-\zeta^2)\left[ -2\zeta(2-\zeta^2) + (3-\zeta^4)\ln\frac{1+\zeta}{1-\zeta}\right],
\end{equation}
where $r_e \approx 2.8 \times 10^{-15}$ m is the classical electron radius and $\zeta \equiv \sqrt{1-1/s}$, with 
\begin{equation}
s = \frac{\epsilon_{\gamma_1} \epsilon_{\gamma_2}}{2m_e^2c^4} \left(1-\cos\Phi \right).
\end{equation}  
This cross-section has a threshold, $s > 1$, dictated by the energy conservation, which translates into the following requirement for the energies of colliding photons:
\begin{equation} \label{eq.Cond}
	 \epsilon_{\gamma 1} \epsilon_{\gamma 2} > \frac{2 m_e^2 c^4}{1 - \cos \Phi}.
\end{equation}
In order to recreate the necessary conditions for observing this process, not only high photon densities, but also high photon energies are required.

Two different approaches have been proposed that rely on high-power lasers for overcoming these challenges. One approach is to fire a gamma-ray beam into the high-temperature radiation field of a laser-heated hohlraum~\cite{Nature.8.434436}, whereas the other approach is to actually collide two gamma-ray beams~\cite{PhysRevE.93.013201}. In what follows, we explore the second approach of using two gamma-ray beams, each like the one obtained from the PIC simulation in the previous section of the paper. These beams are particularly suited for this approach, since they have a large concentration of energetic photons~\cite{PhysRevLett.116.185003}.

%taking each one to have the spectra and angular profile of the one obtained in the previous section of the paper from the PIC simulation.

We consider two scenarios where these two beams collide at an angle of $\Phi = 15^\circ$ and at an angle of $\Phi = 90^\circ$. In both cases, the collisions occur roughly 5 mm away from the laser-irradiated targets that produced the gamma-ray beams. The collision location is deliberately removed from the original targets in order to reduce possible interference from particles and photons other than the gamma-rays of interest. Because of the large distance to the interaction volume compared to the source size itself, we only consider photons that were emitted into an opening angle of $3^\circ$ from the laser axis for each beam. Our goal is to evaluate the pair production rate under such restrictive conditions, which can pave the way for designing experimental setups in the future. 

In the case of the $\Phi = 15^\circ$ collision angle, only photons that satisfy the condition 
\begin{equation}
	\frac{\epsilon_{\gamma 1} \epsilon_{\gamma 2}}{m_e^2 c^4} > 59
\end{equation}
will contribute to the pair-creation. This condition follows directly from Eq.~(\ref{eq.Cond}). The cross-section peaks at 
\begin{equation} \label{EQ-19}
	\frac{\epsilon_{\gamma 1} \epsilon_{\gamma 2}}{m_e^2 c^4} \approx 100
\end{equation}
with a value of 
\begin{equation}
\sigma_{\gamma \gamma}^{\max} \approx 0.75 \pi r_e^2
\end{equation}
and then slowly decreases for higher values of $\epsilon_{\gamma 1} \epsilon_{\gamma 2}$. These estimates indicate that multi-MeV photons will be the major contributors to the pair-production. It is worth pointing out that the photon numbers produced by the source considered in Sec.~\ref{Sec-3} dramatically increase as we decrease the photon energies. Guided by these considerations and the intention of making the pair production simulation feasible, we retain only photons with energies above 1 MeV in the two beams.

\begin{figure} [H]
   \begin{center}
      \includegraphics[width=\columnwidth,trim={0.2cm 5cm 0 5cm},clip]{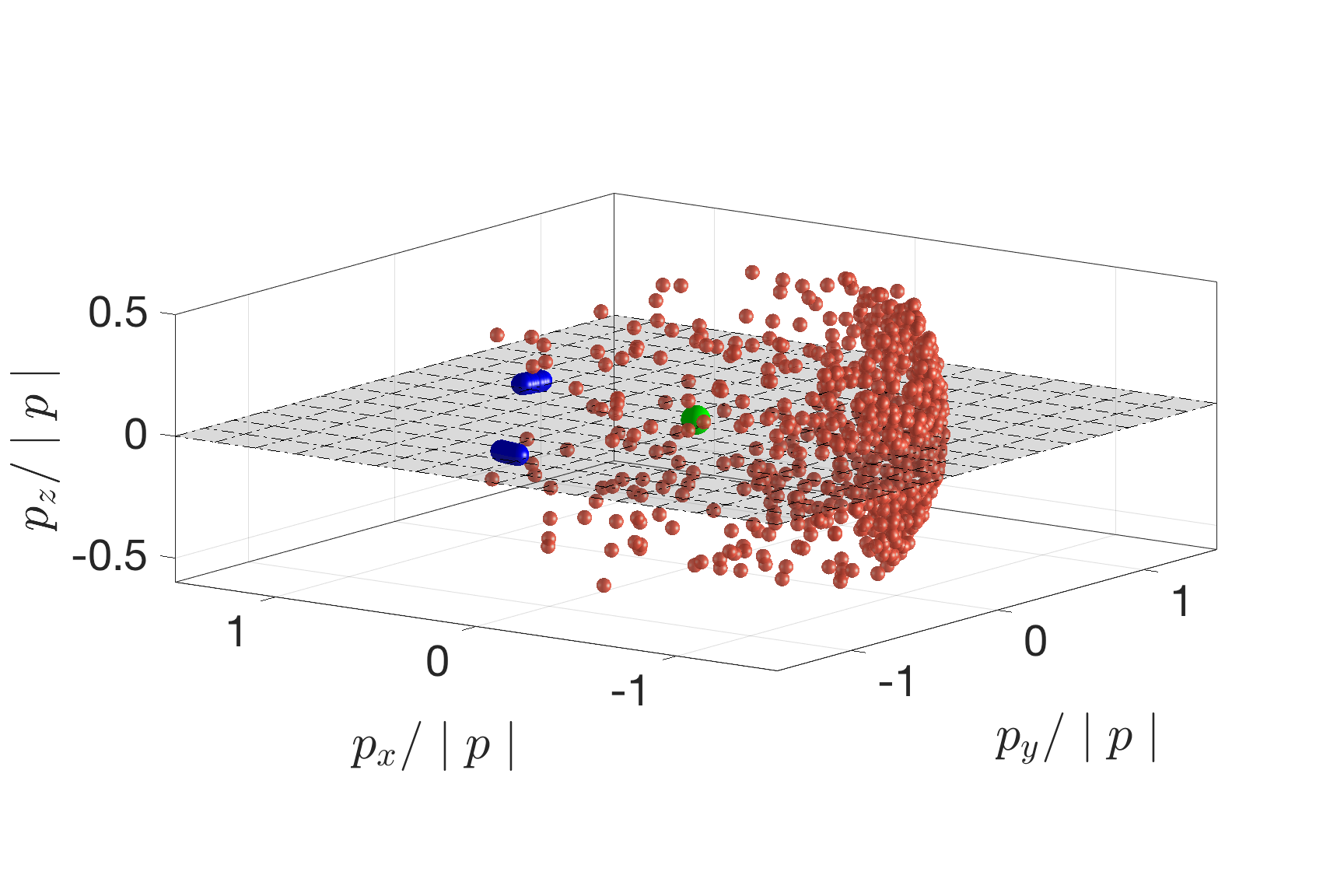} \\
      \includegraphics[width=\columnwidth,,trim={0.2cm 5cm 0 5cm},clip]{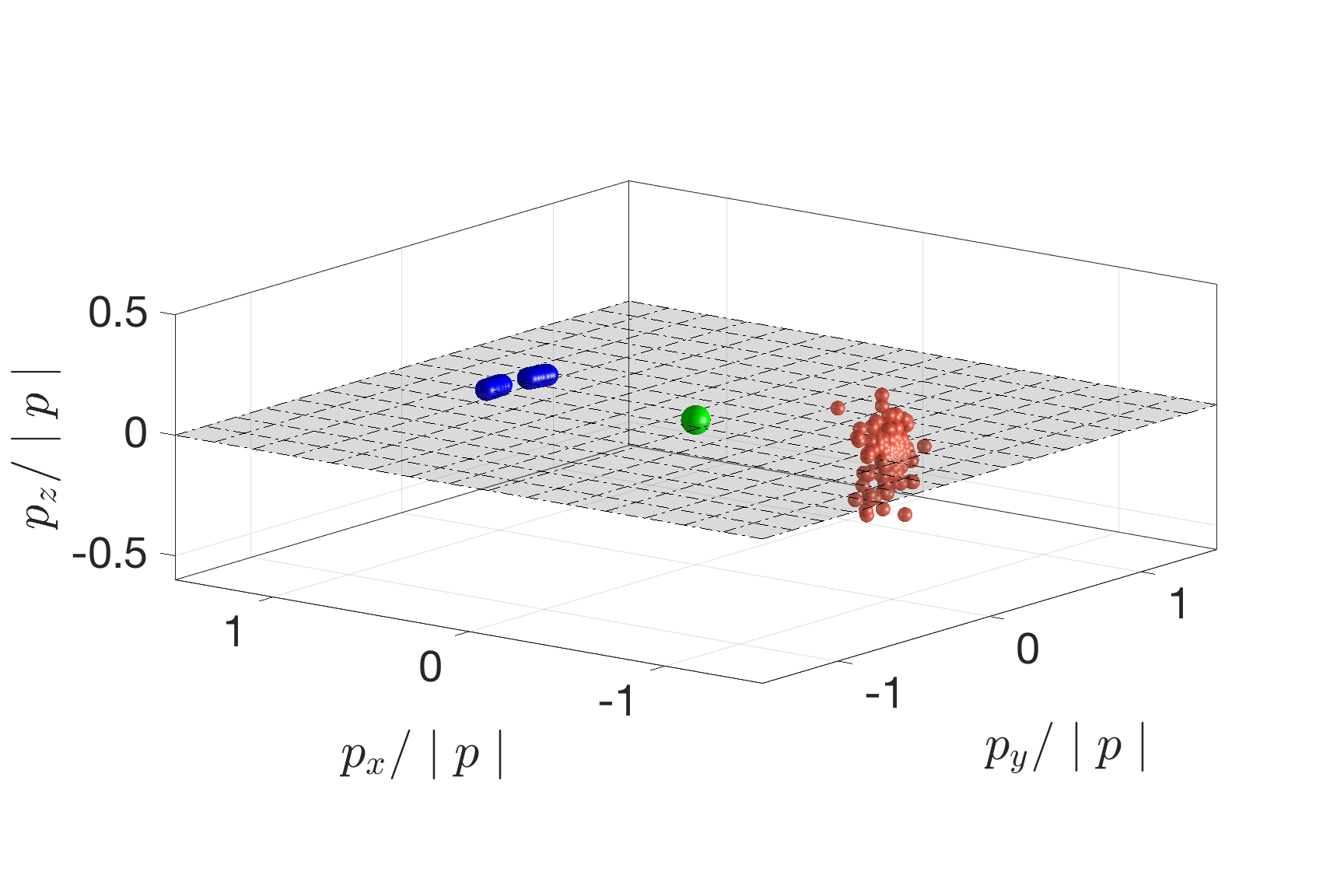}
      \caption{\label{fig:PosLasMomenta_90} Scatterplot of positrons (represented by macro-particles) based on the direction of their momentum. The blue markers represent photons in the incoming gamma-ray beams, with the momentum inverted (with respect to the origin shown by a green marker) to prevent an overlap with positron markers. The collision angles for the beams are $\Phi = 90^\circ$ (upper) and $\Phi = 15^\circ$ (lower).}
      \end{center}
\end{figure}

Simulating the photon-photon collisions in two colliding photon beams presents a serious computational challenge. Even after applying the criteria described above, we are left with over $10^5$ macro-particles representing the photon spectrum shown in Fig.~\ref{fig:PhotAnglHist}. This would then require at least $10^{10}$ binary collision tests in order to check for all of the macro-particles whether or not collisions have occurred. Even for the simplest collision tests, this would lead to unacceptable computation times.

In order to overcome this difficulty, we use the TrI-LEns code~\cite{JANSEN2018582} that was specifically designed for swift collision detection amongst large numbers of particles. Unlike a PIC code, TrI-LEns does not use a mesh of cells; particles move freely in space and are managed in a tree-hierarchy. To dramatically reduce the number of binary collision tests, TrI-LEns uses a modern collision detection algorithm based on bounding volumes that operates with little computational effort and without sacrificing accuracy.

We initialize the photon collision simulation by importing photon macro-particle data generated during the PIC simulation. In TrI-LEns, each macro-particle is interpreted as a rectangular prism (hereafter ``box'') uniformly filled with photons. To translate the PIC data to this format, we uniformly subdivide each cell defined in the PIC code by the number of macro-particles assigned to each cell to make one box for each macroparticle. In order to perform a 3D simulation, we assign to each box a height of 1 $\mu$m, which is roughly the transverse size of the channel in which the photons are generated. The subdivision is only performed in the plane of the PIC simulation, and all photons in one box have the same momentum as the original macro-particle. The described procedure of distributing photons in space is essential to provide the necessary input for the collision detection in the simulation.

\begin{figure} [H]
   \begin{center}
      \includegraphics[width=\columnwidth]{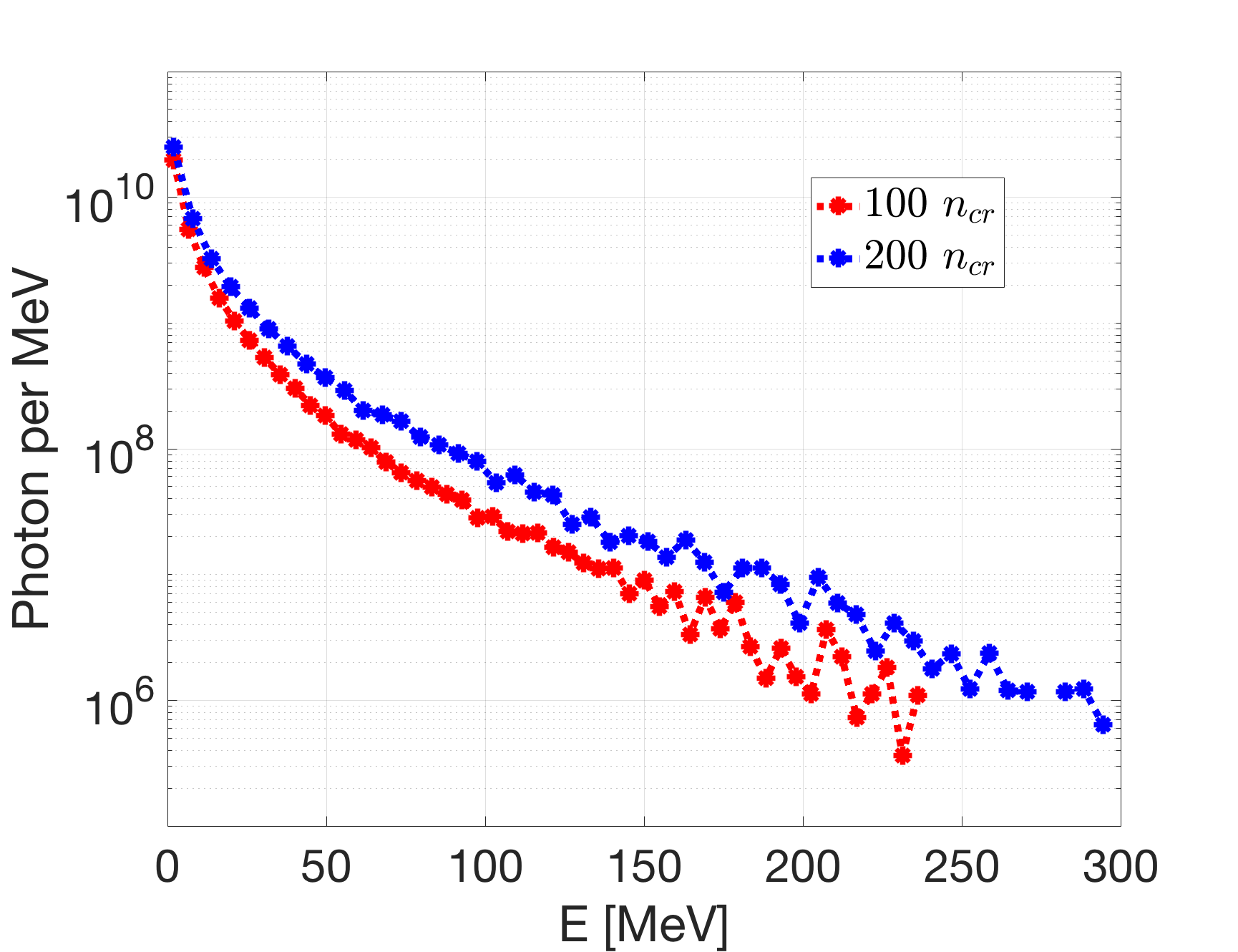} 
      \caption{\label{fig:PhotEnergyHist} Energy distribution of photons in the two colliding photon beams prior to the collision. The upper curve is generated by a source with a bulk target density of $n_e = 200 n_{cr}$. The lower curve is generated by a source with a bulk target density of $n_e = 100 n_{cr}$. The corresponding angle resolved distribution is shown in Fig.~\ref{fig:PhotAnglHist}.}
      \end{center}
\end{figure}

Once the photon boxes have been initialized in space with their appropriate densities and momenta, TrI-LEns can evolve their motion forward in time to check for collisions. Time steps consist of testing for collisions and then pushing the virtual photon boxes. The code currently does not involve any propagating fields and has no mesh, so there is no stability condition that typically severely limits the time step in PIC simulations. This allows for a significant speed up of our calculations. The only aspect that is impacted by the time-step is the chronological order of the collisions. Even with our code, it is computationally demanding to recover the exact chronological order of all of the photon collisions~\cite{JANSEN2018582}.

The collision algorithm implemented in TrI-LEns first checks for the overlap between the rectangular boxes filled with photons. Once an overlap is detected between two boxes, the code calculates an actual interaction volume between the photons. It then uses the cross-section given by Eq. (\ref{BWcross}) and the interaction volume to calculate how many electron-positron pairs are produced in total by the colliding photons from the two boxes. Note that the cross-section is the same for all of the collisions in one such event, because all photons in one box have the same momentum.

An electron and a positron are emitted from each of these pair production events as two macro-particles. The emission is calculated in a frame of reference where the two boxes of photons collide head-on. The direction of emission in this frame of reference is isotropic and it is therefore chosen randomly, but the momenta of the emitted macro-particles are constrained by the energy and momenta conservation requirements. Note that the angular distribution of the emitted pairs is no longer isotropic when transformed back to the laboratory frame of reference if the photon boxes collide at an angle. Once two boxes of photons have collided, the code reduces the number of photons in each box by the number of emitted electron-positron pairs to account for the photon annihilation. The photons are again assumed to be uniformly distributed, so the code effectively reduces the photon density in each box. The pair creation in our algorithm has a threshold, with a pair being produced in a collision of two virtual boxes only if the calculated number is greater than unity. 

Figure~\ref{fig:PosLasMomenta_90} shows  the positron yields from the collisions between two gamma-ray beams that we simulated. The upper panel is for a $\Phi = 90^\circ$ collision between the beams that produces a total of 1153 positrons, whereas the lower panel is for a $\Phi = 15^\circ$ collision that results in 173 positrons. All incoming photons are moving in the $(x,y)$-plane shown with color in both panels. In these scatter plots, the markers are macro-particles representing generated positrons. Since the events are relatively rare, one macro-particle roughly corresponds to one positron. 

It must be pointed out that two different photon sources are required to simulate the collision. Taking two identical sources would cause for an artificially high number of photons to collide, which is a geometrical effect. Instead of introducing additional randomization, we decided to use two sources that were produced by targets with different bulk densities. One source is from the exact setup described in the previous section that had a bulk density of $n_e = 100 n_{cr}$. The second source was instead calculated using a bulk density of $n_e = 200 n_{cr}$ but with the same parameters for the laser pulse. The photon spectra that were used for the photon collision simulations are shown in Fig.~\ref{fig:PhotEnergyHist}.

\begin{figure} [H]
   \begin{center}
      \includegraphics[width=\columnwidth]{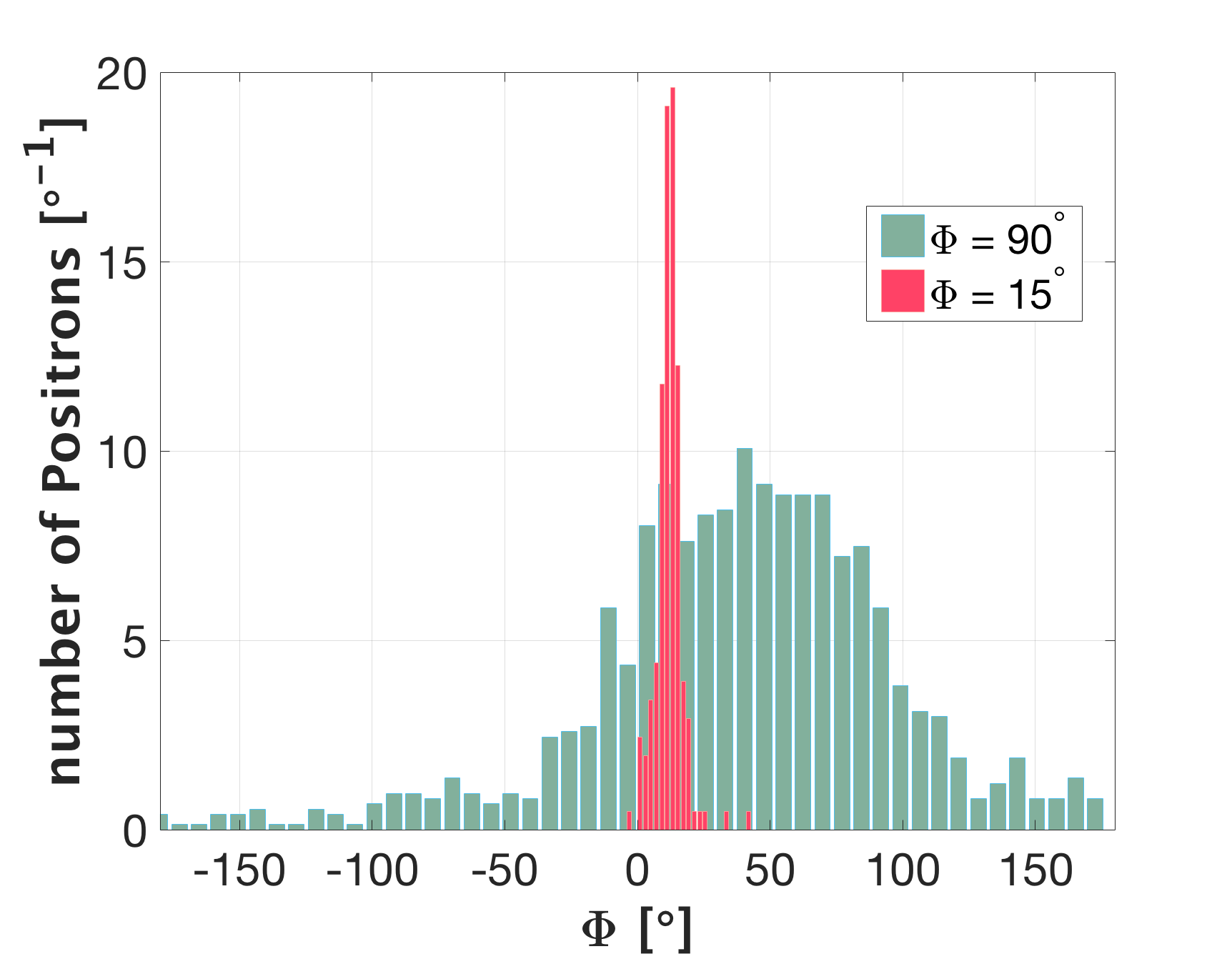} 
      \caption{\label{fig:PosAngleHist} Histogram of the azimuthal angle $\phi$ representing the direction of the positron momentum in the plane of the colliding photon beams. The narrow distribution is for $\Phi = 15^\circ$ (red) and the wider distribution is for  $\Phi = 90^\circ$ (green).}
      \end{center}
\end{figure}

The difference in the photon spectra causes the emitted positron beam to be slightly asymmetric. Angular distributions of the positrons in the plane of the colliding photon beams are plotted in Fig.~\ref{fig:PosAngleHist} for the two collision angles that we have considered. The colliding photon beams are centered around $\phi = 0^{\circ}$, so the resulting positron beams would be symmetric for photon beams with the same energy distribution.

We can then conclude that, despite the fact that the interaction region is significantly removed from the laser-irradiated targets that generate the photons, we are able to generate on the order of $10^3$ pairs. The collimation of the generated positrons can be improved by reducing the collision angle, as clearly shown in the lower panel of Fig.~\ref{fig:PosLasMomenta_90}. However, improved collimation comes at the expense of the positron yield. It remains to be determined whether such a trade-off is beneficial in the context of the schematic setup shown in Fig.~\ref{fig:BWexperiment}.

%*******************************************************

\section{Summary and Conclusions} \label{Sec-5}

We have examined the role of a plasma magnetic field driven in a classically over-critical plasma by a high-intensity laser pulse. Specifically, PIC simulations demonstrate the potential of such setups to exploit the phenomenon of relativistic transparency, which enables a laser-driven current capable of generating magnetic fields of unprecedented strength. This magnetic field, in turn, is shown to significantly enhance the emission rates of ultra-relativistic electrons traversing the field. An analysis of various kinds of electron motion in a laser field, including the direct laser acceleration mechanism, highlight the non-trivial change that a background magnetic field induces in electron motion and radiation rates. PIC simulations corroborate this analysis by showing that a classically over-critical channel embedded in a near-critical bulk target facilitates the emission of large quantities of multi-MeV gamma-rays into a narrow angular cone, promoted by the large magnetic field at the channel edges. 

A host of potential applications arises from this type of gamma-ray generation capability, but we focused here on the exciting prospect of observing the linear Breit-Wheeler process of pair production. By colliding two such dense, collimated gamma-ray beams that can be created through our laser-target setup, the TrI-LEns tree code demonstrates positron production of varying yield (up to $\sim 10^3$) and directionality based on the angle of incidence. Therefore, this technique of gamma-ray production opens a new door into fundamental physics research and gamma-ray applications that can be realized with today's technology.

\section*{Acknowledgements}
This research was supported by the National Science Foundation under Grant No. 1632777 and the US Air Force project AFOSR No. FA9550-17-1-0382. Simulations were performed using the EPOCH code (developed under UK EPSRC Grants No. EP/G054940/1, No. EP/G055165/1, and No. EP/G056803/1) using HPC resources provided by the TACC at the University of Texas and the Comet cluster at the SDSC at the University of California at San Diego.

\section*{References}
\bibliography{channel}
\bibliographystyle{unsrt}

\end{document}